\documentclass[doublecol]{epl2} 
\usepackage{amsmath}
\usepackage{amssymb}
\usepackage{siunitx}
\usepackage{hyperref}
\title{Exotic Quantum Phenomena in Twisted Bilayer Graphene}
\author{G. Feraco \inst{1}, W. Boubaker \inst{1},  P. Rudolf \inst{1}, A. Grubi\v si\' c-\v Cabo \inst{1}}
\institute{
    \inst{1} Zernike Institute of Advanced Materials University of Groningen 
}
\abstract{
      Bilayer graphene twisted at the angle of about $\SI{1.1}{\degree}$, better known as magic angle, exhibits ultra-flat moiré superlattice bands that are a source of highly-tunable, exotic quantum phenomena. Such phenomena, like superconductivity, correlated Mott-like insulating states and orbital ferromagnetism are driven by strong-correlation physics that defies classical theories. The inadequacy of such classical models and the lack of theoretical understanding of the recently observed exotic phenomena calls for revisiting the theory behind the material system and associating it with the observed behaviour. This article reviews the physics behind twisted bilayer graphene, focusing primarily on moiré physics and the importance of electronic (flat) band structure. In addition, this paper provides a brief overview of the emerging phenomena of correlated insulating states, superconductivity and orbital ferromagnetism. Finally, the most recent developments in controlling the interaction-driven states and tuning the electronic interactions are presented.
}
\begin{document}
\maketitle
Bilayer graphene (BG) is a van der Waals material that consists of two graphene layers held together by van der Waals interaction \cite{2018ja}. In its most common form, Bernal- or AB-stacking, BG is a gapless semiconductor, and similar to the case of single layer graphene (SLG) \cite{Geim2004} this is limiting its uses for switching applications. A band gap can be introduced in BG by applying an external electric field in the direction normal to the surface due to the breaking of inversion symmetry in this direction \cite{electronic-properties-graphene}, or by replacing carbon with other atoms, such as nitrogen \cite{NG} or boron \cite{BG}. An alternative approach is to use altogether different materials to develop similar properties, such as transition metal dichalcogenides (TMDs) \cite{thesis-Cao}. A more novel way of introducing a band gap in BG is by twisting one of the layers relative to the other and creating a twisted structure. In certain cases the interaction between the twisted layers becomes so strong that it leads to an emergence of exotic phenomena such as superconductivity, correlated insulating states and magnetism, among the rest \cite{marvels}, revealing the incompetence of old models, such as the Hubbard model, to explain the emergent behaviour in these materials. These aforementioned discoveries opened up a whole new discipline in physics, nowadays referred to as moiré physics and consequently it started the field of twistronics.
Twistronics became widely popular following the discovery of unconventional superconductivity and correlated insulating states in BG twisted at a magic angle of $\SI{1.1}{\degree}$ by the group of Pablo Jarillo-Herrero in 2018 \cite{Pablo-2018, Pablo-2018(2)}. Remarkably, it turned out that twisted bilayer systems can be used as a single-material platform to study strongly correlated physics, with different phases being reachable simply by using electrostatic gating \cite{Pablo-2018, Review-Eva}.

\subsection{Moiré physics}
\begin{figure*}[t!]
    \includegraphics[width=1\textwidth]{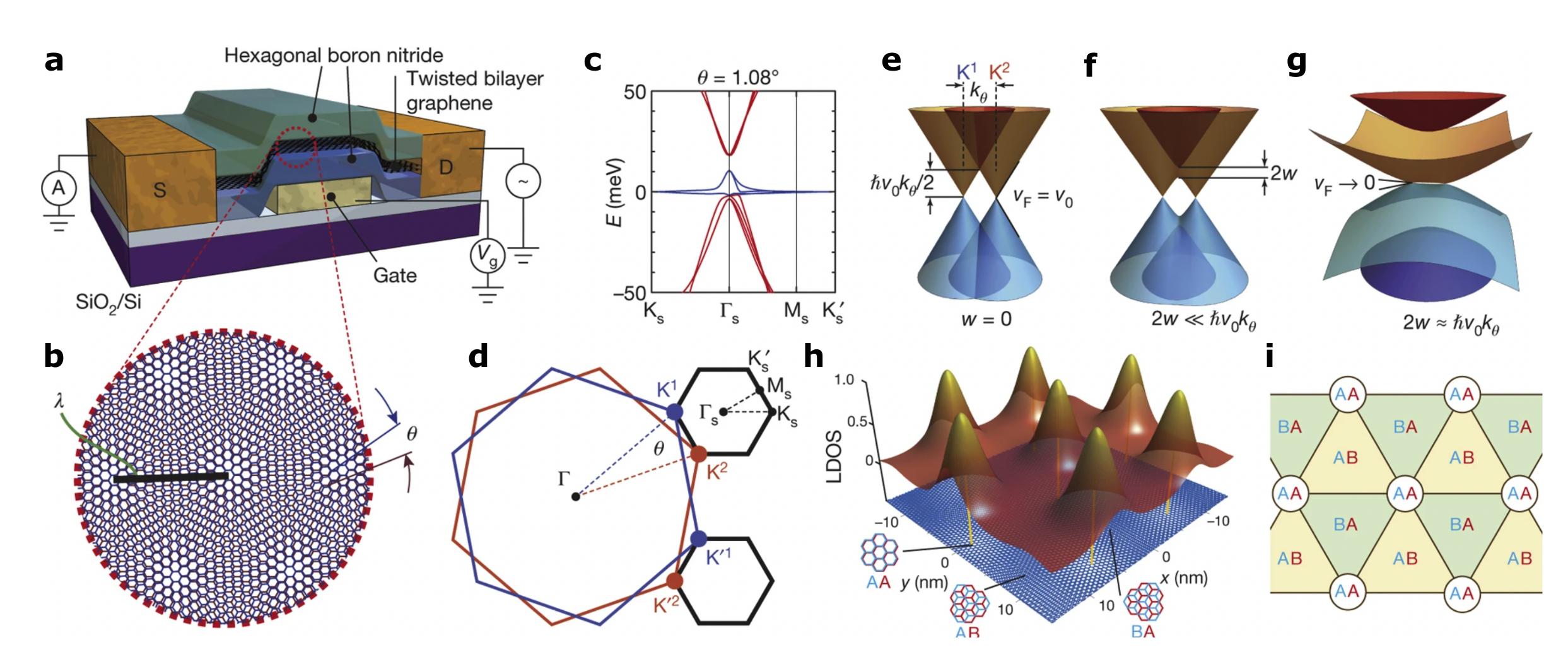}
    \caption{\textbf{a.} Sketch of a transistor setup used to control the charge density in the twisted bilayer graphene (TBG) structure, encapsulated between flakes of hBN.
    \textbf{b.} Moiré pattern in TBG characterised by a large moiré superlattice, which periodicity is indicated by the black bar.
    \textbf{c.} Band diagram of TBG at an angle of $\theta=\SI{1.08}{\degree}$. 
    \textbf{d.} The \text{mini} Brillouin zone of the moiré pattern corresponds to the difference between the two wavevectors of the twisted layers. 
    \textbf{e-g.} Schematic representation of the effect of interlayer coupling and hybridisation, when the interlayer coupling is: $w = 0$ (\textbf{e}), $2w\ll \hbar v_0 k_\theta$ (\textbf{f}) and $2w \approx \hbar v_0 k_\theta$ (\textbf{g}), where $v_0$ is the Fermi velocity of single layer graphene ($v_0 = \SI{e6}{\meter\per\second}$). \textbf{h.} Normalized local density of states at $\theta = \SI{1.08}{\degree}$.
    \textbf{i.} Simplified schematic representation of the spatial stacking order in TBG (top view). Reprinted with permission from ref. \cite{Pablo-2018(2)}.}
    \label{fig:setup}
\end{figure*}

In layered van der Waals systems the weak van der Waals forces between the layers allow for the relative rotation of layers, which is not possible in other types of systems. 
In the simplest case of two van der Waals layers coming from the same material, rotation of layers with respect to one another gives rise to a so called moiré pattern, as illustrated in Figure \ref{fig:setup}b.  These moiré patterns manifest as long-range interference patterns in atom densities, and a simple change of the twist angle induces a significant transformation in the moiré pattern. Consequently, a moiré pattern constructs a periodic superlattice, with a wavelength ($\lambda_M$) substantially exceeding the crystal unit cell, and it is inversely proportional to the twist angle ($\theta$), as shown in equation \ref{eq:wavelength_superlattice},\cite{Pablo-2018(2), Review-Eva}:
\begin{equation}
    \lambda_{M} = \frac{2 \pi}{\sin{\left(\pi/3\right)}K_M} = \frac{a}{2\sin{\left(\theta/2\right)}},
    \label{eq:wavelength_superlattice}
\end{equation}
where $a$ is the lattice constant of graphene ($a = \SI{2.46}{\angstrom}$) and $K_M$ is the moiré wavevector  defined as the difference between the two lattice wave vectors (\textbf{k$_1$} and \textbf{k$_2$}) and, therefore, in the case of BG has a hexagonal \textit{mini} Brillouin zone in momentum space corresponding to the moiré superlattice arising from the superposing the Brillouin zones of the two graphene layers (see Figure \ref{fig:setup}d). The smallest wave vector difference between the two reciprocal lattice vectors is given by $\Delta\mathbf{k} = \mathbf{k}_1-\mathbf{k}_2$ and for geometric reasons it follows:
\begin{equation}
    K_M = |\Delta\mathbf{k}| = |\mathbf{k}_1|\cdot 2\sin{\left(\frac{\theta}{2}\right)}.
    \label{eq:wavevector}
\end{equation}
Equation \ref{eq:wavevector} shows a relation between the wavevector of the original graphene lattice and the twist angle. For small twist angles, equation \ref{eq:wavelength_superlattice} can be simplified further into \ref{eq:wavelength2} \cite{thesis-Cao}:
\begin{equation}
    \lambda_M \approx \frac{a}{\theta}.
    \label{eq:wavelength2}
\end{equation}
As can be seen from the equation \ref{eq:wavelength2}, the smaller the twist angle the larger the wavelength of the moiré pattern. 
In the case of BG, the number of charge carriers required to fill in the moiré unit cell is four and the corresponding density $n_s$ is the charge density required to fill the bands up to the insulating gap \cite{Pablo-2018, Pablo-2018(2), thesis-Cao}. Through this, it is possible to define a band-filling factor $\nu$ as the number of charge carriers per moiré unit cell. Charge neutrality corresponds to $\nu = 0$, while $\nu = \pm 4$ corresponds to bads completely filled ($+$) or completely empty ($-$). Intermediate values $\nu = \pm 1, \pm 2, \pm 3$ correspond to a partial filling by one, two or three electrons ($+$) or holes ($-$) per moiré unit cell \cite{Review-Eva}.
A precise control of the charge density can be achieved using a field effect transistor as shown in Figure \ref{fig:setup}a. 
The device consists of a twisted bilayer graphene which is encapsulated between flakes of insulating hexagonal boron nitride (hBN) and is connected to a bottom gate and source/drain contacts. The hBN layers act as an insulating dielectric for electrostatic gating and a protection layer for the graphene bilayers \cite{Pablo-2018(2)}.
While both SLG and BG do not possess a band gap, the electronic band structure of BG is nevertheless significantly different when it comes to its dispersion relation. Namely, BG has quadratic dispersion and massive carriers, in stark contrast to linear dispersion and massless carriers in SLG. The situation is significantly different for twisted BG (TBG). Twisting the layers in real space causes rotation of the reciprocal lattice in momentum space, thereby creating a shift between the electronic band structures of the two layers as shown in Figure \ref{fig:setup}b-h. The physics in TBG mainly depends on one single dimensionless parameter $\alpha$ \cite{thesis-Cao}, which relates the interlayer coupling $w$, also called hopping parameter, to the SLG Fermi velocity $v_0$, the twist angle $\theta$ and the wave vector $k_{\theta}$ \cite{Bistritzer-MacDonald}:
\begin{equation}
    \alpha = \frac{w}{\hbar v_0 k_{\theta} \theta}.
    \label{eq:parameter-alpha}
\end{equation}
In the case of weak interactions between the two layers (large twist angles), the interlayer coupling $w$ is insignificant ($w \approx 0$) and the band structure of individual single layers cross without causing any considerable changes to the electronic structure, as seen in Figure \ref{fig:setup}e. In this regime, the interlayer coupling could be considered a mere perturbation to the Dirac cones. In the case of TBG, the interlayer coupling $w$ becomes significant for small angles ($2w \ll \hbar v_0 k_{\theta}$) and leads to hybridization, which opens a band gap in the band structure, as shown in Figure \ref{fig:setup}f. In this case the coupling can no longer be considered a simple perturbation as it induces significant changes in the electronic band structure.
Further increase in the interlayer coupling, $2w \approx$ $\hbar v_0 k_{\theta}$, causes the down shift in the lower hybridised state, which crosses zero energy. 
As a consequence, the band gap can be tuned through the twist angle $\theta$.
At specific angles, known as the \textit{magic angles}, the bands become completely flat; this phenomenon is known as the \textit{Flat Band} condition, which was termed by Bistritzer and MacDonald \cite{Bistritzer-MacDonald} and experimentally visualized by Lisi \textit{et al.} \cite{twist-angle-disorder2} and Utama \textit{et al.} \cite{utama2021visualization} using angle-resolved photoemission spectroscopy with sub-micron spatial resolution. 
Flat bands in momentum space correspond to strong electron localisation in real space. This localisation effect is a direct consequence of the band structure, which depends on the moiré potential. The Fermi velocity of TBG $v_F$, can be determined from the slope dE(\textbf{q})/dq of the energy diagram at the Dirac neutrality points.
The phenomenon of decreasing Fermi velocity with decreasing twist angles is called renormalization of the Fermi velocity. The latter can also be deduced from equation \ref{eq:parameter-alpha}; for larger values of $\alpha$, corresponding to small twist angles, the interlayer coupling $w$ increases and the Fermi velocity $v_F$ decreases. 
The Fermi velocity $v_F$ can, in a first order, be related to the SLG Fermi velocity $v_0$ via equation \ref{eq:fermi-velocity} \cite{Bistritzer-MacDonald}:
\begin{equation}
    v_F(\alpha) = \Bigg| \frac{\partial E_K \left(\mathbf{q}\right)}{\partial\mathbf{q}} \Bigg| = \frac{1-3\alpha^2}{1+6\alpha^2} \cdot v_0,
    \label{eq:fermi-velocity}
\end{equation}
where $E_K$ is the energy at which the Dirac cones from the two layers intersect. The Fermi velocity $v_F(\alpha)$ goes to zero when $\alpha = \frac{1}{\sqrt{3}}$, which for BG corresponds to $\theta_M^1 = \SI{1.1}{\degree}$, defining the first magic angle. Higher-order magic angles exist, but unlike the first magic angle $\theta_M^1$, are more dependent on model parameters and, therefore, less robust to perturbation and less stable \cite{higher-order-magic-angles}.
Besides the angle between the layers, it is worth mentioning that also the alignment with the substrate plays a role in changing the electronic properties of the material. The substrate most often employed for devices is hBN, and it was observed that the degree of alignment of TBG and hBN has strong effect on the appearance of the unconventional superconducting and correlated insulating states (TBG and hBN misaligned) or their absence (TBG and hBN aligned) \cite{Serlin2020}.
Recent studies also show gaped phases at fractional filling, when partially alignment occurs due to the presence of a super-moiré super lattice arising from the interaction with the hBN substrate/capping layer \cite{Wong2023}.
The suppression of superconducting phase also occurs by employing a different substrate, such as WS$_2$. In this case, when the magic angle TBG (MATBG) and the substrate are misaligned the spin orbit coupling can have strong influence on the elctronic properties of the material, which show robust anomalous Hall effect (AHE) at quarter and half-filling \cite{Haddad2023}. Hence, the choice of the substrate can affect the properties of the material and can be exploited as a substrate engineering for tuning topological properties in MATBG \cite{Lin2022}.
The alignment of the substrate is only one of the parameters that can be used to effectively tune the electronic properties of TBG, flat bands can appear in TBG at non-magic angles due to strain \cite{Zhang2022}, which can also locally modify the moiré pattern \cite{Dey2023}, or doping, that can be used to tune correlation effect by inversion symmetry breaking, opening gaps at the Dirac points \cite{Dale2023}.
Additionally, the presence of Coulomb impurities affect the shape of the flat bands; the deformation strongly depends on the position of the charged defects in the moiré unit cells suggesting that a defect engineering can be exploited to control electronic phases of TBG \cite{Ramzan2023}.
Finally, strong interlayer hybridisation and moiré flat bands are not only limited to TBG but have also been observed in twisted TMDs such as bilayer WSe$_2$ \cite{WSe2}, TMD hetero-bilayers \cite{TMD-hetero2}, twisted bilayer-bilayer graphene \cite{bilayer-bilayer-graphene1}, double twisted bilayer graphene \cite{Ding2023} and others.
Recent investigations highlight that also more complex structures as monolayer-trilayer graphene heterostructure \cite{Zhou2023} or twisted double trilayer graphene \cite{Perrin2024} show the presence of correlated states and can be employed as a highly tunable platform for further exploration of optoelectronic properties. 
\subsection{Correlated insulating states}
 \begin{figure*}[h!]
 \centering
    \includegraphics[width=0.7\textwidth]{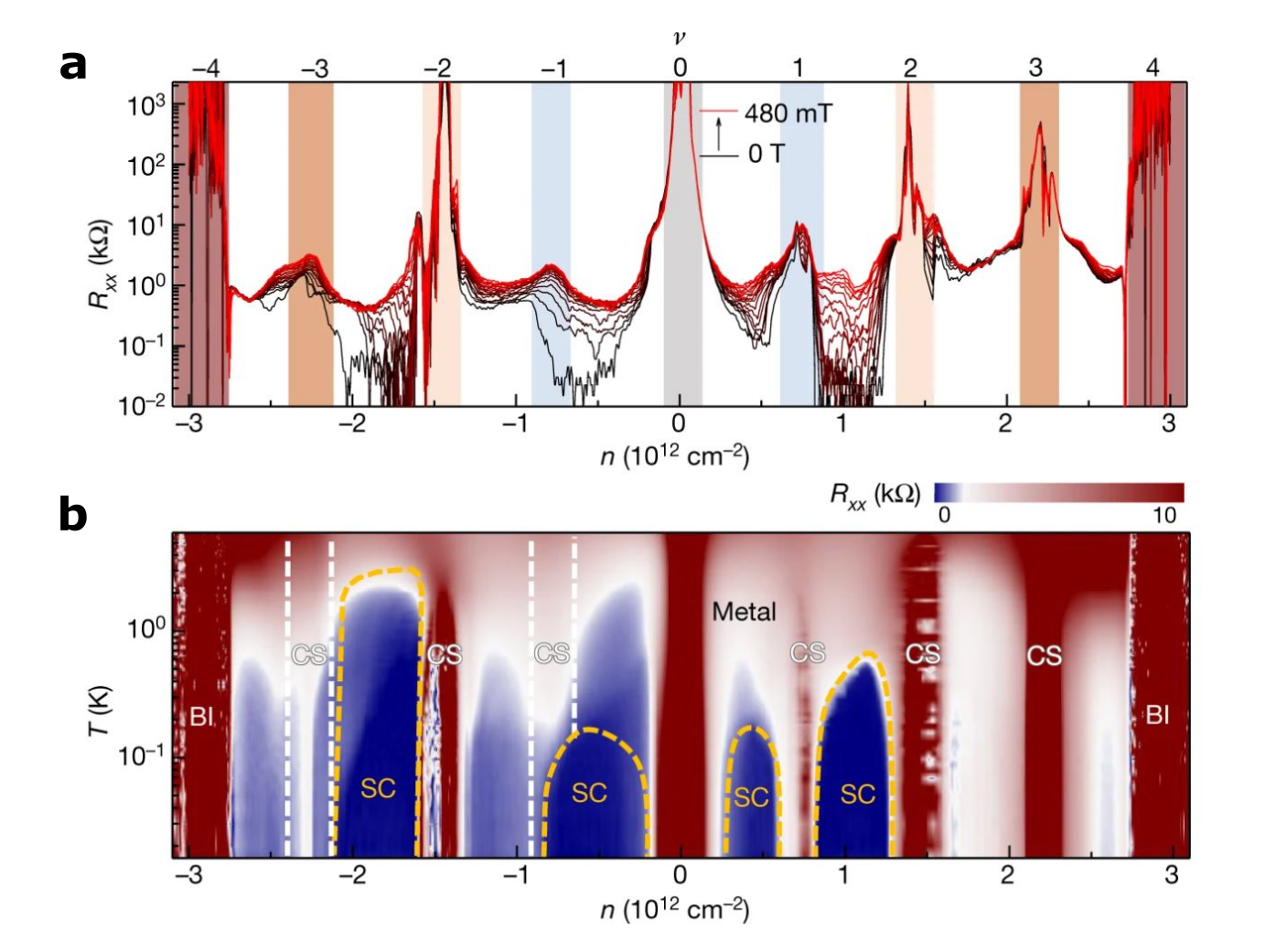}
    \caption{Correlated insulating states and superconducting domes in magic angle twisted bilayer graphene. \textbf{a.} Longitudinal resistance of TBG ($\theta = \SI{1.10}{\degree}$) at $T = \SI{16}{m\kelvin}$ as a function of the charge density $n$,  at different perpendicular magnetic fields from 0 T (black trace) to 480 mT (red trace), controlled via electrostatic gating.
    \textbf{b.} Phase diagram showing longitudinal resistance against carrier density and temperature. Different phases are marked as metal, band insulator (BI), correlated state (CS) and superconducting state (SC).
    Reproduced with permission from ref. \cite{Lu2019}.
    }
    \label{fig:insulating-states}
\end{figure*}
As discussed in the previous section, twisting two layers of graphene can create flat bands in the system at certain angles termed magic angles. The narrow bandwidth ($W$) of the flat bands has direct consequences on the material's properties, which, most naturally, depend on band filling. Band filling is determined by the number of charges occupying available states near the Fermi energy level. This charge density in the system can be precisely controlled via electrostatic gating and it is employed in BG field effect transistors. Figure \ref{fig:insulating-states} shows the density-dependent longitudinal resistance ($R_{xx}$) at different perpendicular magnetic fields from $0$ $T$ (black trace) to $480$ $mT$ (red trace) measured at $T = \SI{16}{m\kelvin}$ for a device with a twist angle $\theta = \SI{1.10}{\degree}$.
Strong resistance peaks observed at $\nu = \pm4$ are a direct result of the presence of a band gap above and below the flat band of the superlattice. Moreover, interaction-induced resistance peaks at all non-zero integer fillings of the moiré bands $(\nu = \pm1,\pm2,\pm3)$ are also present.
These results are in accordance with ref. \cite{Pablo-2018(2)} where it was observed that transitions occur near half-filling of the superlattice bands ($\mu = \pm 2$), \textit{i.e.} near $n = \pm (n_s/2) = \pm (1.4 \pm 0.1) \times 10^{12}$ $\SI{}{\per\centi\meter\squared}$. At this value, the system undergoes a metal-to-insulator transition (MIT) as the conductivity decreases abruptly, in contradiction with the conventional band theories which postulate that solids with half-filled valence bands should be electrically conducting.
Nevertheless, these theories fail to consider electron-electron interactions, which can become significant and completely change the system's behaviour. In fact, at low temperatures, electron-electron interactions, determined by a delicate interplay of screening processes taking place in the material \cite{Pizarro2019}, can cause a classically conductive material to become insulating; as in the case for Mott insulators, a class of insulating materials in which the valence band is half-filled. In Mott insulators each available state is occupied by one electron (obeying Hund's rule), and the Fermi energy levels lie within the band. Classically, such a system is expected to show metallic behaviour, but due to strong Coulomb repulsion ($U$) between the electron charges, conduction is suppressed, and a band gap is introduced within bands of the same orbital character, which explains the appearance of an MIT. Essentially, Mott transitions result from the interplay between the Coulomb repulsion in the system and the electronic bandwidth $W$, which is a measure of the degree of localisation. Sufficiently strong Coulomb repulsion promotes the localisation of formerly conducting charges, thereby lowering the system's total energy and reducing the bandwidth by forming a gap within the band \cite{Mott1}. 
At temperatures above $\SI{4}{\kelvin}$, the TBG system behaves as a traditional metallic conductor, however, at temperatures below $\SI{4}{\kelvin}$, MATBG displays Mott-like insulating behaviour at half-filling of the bands ($\mu = 2$), which can be ascribed to electron-electron interactions, as discovered by the group of Pablo Jarillo-Herrero in 2018 \cite{Pablo-2018(2)}. The interaction-induced insulating states persist as long as Coulomb repulsion $U$ dominates with respect to the bandwidth $W$, \textit{i.e.} $U/W>1$. This is valid around the magic angle, where the bands are sufficiently flat, \textit{i.e.} they have narrow bandwidth.
Based on the geometrical moiré pattern illustrated in Figure \ref{fig:setup}f, a Hubbard model on a triangular lattice can be considered to estimate the Coulomb energy $U$ at each moiré site, where each site coincides with an AA-stacking (atoms in each graphene layer located directly on top of one another) region in the twisted bilayer. The Coulomb energy (U) can be expressed as follows:
\begin{equation}
    U = \frac{e^2}{4\pi\epsilon d} = \frac{e^2 \theta}{4\pi \epsilon \kappa a},
    \vspace{0.2cm}
\end{equation}
\noindent where $e$ is the electron charge, $\epsilon$ is the effective dielectric constant, $d$ corresponds to the moiré period length scale, \textit{i.e.} $d \approx \lambda_M \approx a/\theta$ for $\theta \ll 1$ (eq. \ref{eq:wavelength2}), and $\kappa = \kappa\left(\epsilon, d(\theta)\right)$ is a constant that depends on the effective dielectric constant and the moiré period length scale \cite{Pablo-2018(2), thesis-Cao}.
A slight temperature increase of a few degrees is already capable of annihilating these insulating states, due to similar energy scales between the thermal energy and the activation energy, which was measured to be $\sim \SI{0.32}{\milli\electronvolt}$ \cite{Pablo-2018(2)}, much smaller than the activation energy of the superlattice band gap of $\sim \SI{40}{\milli\electronvolt}$ \cite{Pablo-2018(2)}.
The suppression of the insulating states occurs at half-filling, even in the presence of a magnetic field. This phenomenon is due to the Zeeman effect supported by the fact that the corresponding Zeeman energy ($g \mu_B B = \SI{0.5}{\milli\electronvolt}$) is of the same magnitude as the activation energy \cite{Pablo-2018(2), thesis-Cao}. The application of an external magnetic field can further polarize the excitations increasing the Zeeman energy, which can overcome the band gap $\Delta$ and promote charge conduction.
In summary, the correlated insulating states are the combined result of the small bandwidth due to the emergence of flat bands near the magic angle and the large Coulomb interactions between electrons at half-filling of the bands.
\subsection{Superconductivity}
The emergence of flat bands near half-filling does not only give rise to correlated insulating states, but also to other unexpected phenomena such as superconductivity \cite{Pablo-2018}. The appearance of superconducting states in MATBG should not be too surprising -- many correlated materials with broken symmetry states, such as copper oxides (also called cuprates), are characterised by correlated Mott insulating states and have shown superconductive states upon doping. At ambient pressure, cuprate superconductors have the highest critical temperatures ($T_c$), below which they become superconducting, reaching $T_c$ above $\SI{100}{\kelvin}$ \cite{high_Tc-superconductors}. 
Density-temperature phase diagram of MATBG in Figure \ref{fig:insulating-states}b, is astonishingly similar to the phase diagram of high-T$_c$ cuprates \cite{high_Tc-superconductors} (not shown here), where temperature is plotted against the doping level, as both systems display superconducting domes centred around a correlated insulating (CI) phase. 
For both TBG and cuprates, the superconducting (SC) domes are more pronounced for hole-doping than electron-doping.
One significant advantage of using TBG for investigation of a superconducting domes is the ability to tune the charge density via electrostatic gating without the need for fabricating new devices for each single charge density, an incredibly tiresome process that is required in the case of cuprates \cite{Review-Eva}. Point in case, the whole phase diagram (Figure \ref{fig:insulating-states}b) can be mapped using a single sample, whereas cuprates require an infinite amount of samples to obtain a diagram of such high resolution. 
In ref. \cite{Pablo-2018(2)} the temperature-dependent IV curves shows a resistive switching behaviour near $\sim 50$ $\SI{}{\nano\ampere}$, a value at which the system becomes superconducting. A small charge density of $(1.2-1.8) \times 10^{12} \, \SI{}{\per\centi\meter\squared}$ is sufficient to induce superconductivity states in MATBG, which is at least one order of magnitude smaller than other materials, such as doped MoS$_2$ ($n = \SI{7e13}{\per\centi\meter\squared}$) and LaAlO$_3$/SrTiO$_3$ interfaces ($n = \SI{1.5e13}{\per\centi\meter\squared}$) implying that the origin of superconductivity in MATBG is most likely not conventional and cannot be described by the BCS theory. In BCS theory, Cooper pairs (pairs of fermions) are formed at sufficiently low temperatures via the weak coupling between electrons and phonons, allowing electrons to become correlated \cite{BCS1, BCS2}. The critical temperature $T_c$ of a BCS superconductor scales exponentially with the density of states $N$ and the electron-phonon coupling strength $\lambda$: $T_c\:\approx \: \exp{\left(1/\lambda N \right)}$. If MATBG obeyed BCS theory, the DOS below and above $n = \pm (n_s/2)$, around which the superconductive states appear, should be significantly larger. However, experimental measurements of the effective mass confirmed that the DOS for $n \lesssim - n_s/2$ is lower \cite{thesis-Cao} than what is expected for BCS theory. Hence, the superconductive states in MATBG hardly obey BCS theory and are much more likely to emerge due to strong electron correlations, although this is still debated. In fact, there are arguments in favor and against strong-coupling physics governing different phases of MATBG diagram. Besides the behavior at half-filling, where the shape of the phase diagram resembles the one of the cuprates pointing out strong coupling emergence, close to charge neutrality, the robust suppression of conductance against temperature and magnetic field indicates the presence of strong coupling \cite{Pizarro2019}. However, the fragility of the insulating states that can be easily destroyed by temperature above 4 K or magnetic fields above $B$ $=$ $5$ $T$ indicates weak coupling. \cite{Pizarro2019} Crucial step in the future studies of TBG would be to distinguish between weak- and strong- coupling scenarios in order to unveil the nature of the different effects and clarify whether these phenomena can occur in other van der Waals materials.
\subsection{Orbital ferromagnetism}
Graphene is a material with significantly low spin-orbit coupling and is, therefore, classified as a non-magnetic material. Nevertheless, in a research led by Sharpe \textit{et al.} \cite{ferromagnetism2019} ferromagnetism was discovered in BG which was twisted at a near-magic angle of $\theta = \SI{1.20}{\degree} \pm \SI{0.01}{\degree}$ and encapsulated between two layers of hBN, with one of the hBN layers aligned with the TBG. 
This system showed unambiguous evidence of ferromagnetism in MATBG in a narrow range of densities at three-quarter filling, at which MATBG appears to be an insulator, most likely due to correlation effects \cite{Pablo-2018(2), tune-interlayer-coupling}. 
While ordinary ferromagnetism results from spin-polarisation at the Fermi level, the ferromagnetic ordering in this system was found to be caused by the alignment of electron orbital motion. This discovery is the first of its kind, as orbital ferromagnetism had never before been demonstrated experimentally.
\begin{figure*}[h]
 \centering
    \includegraphics[width=1\textwidth]{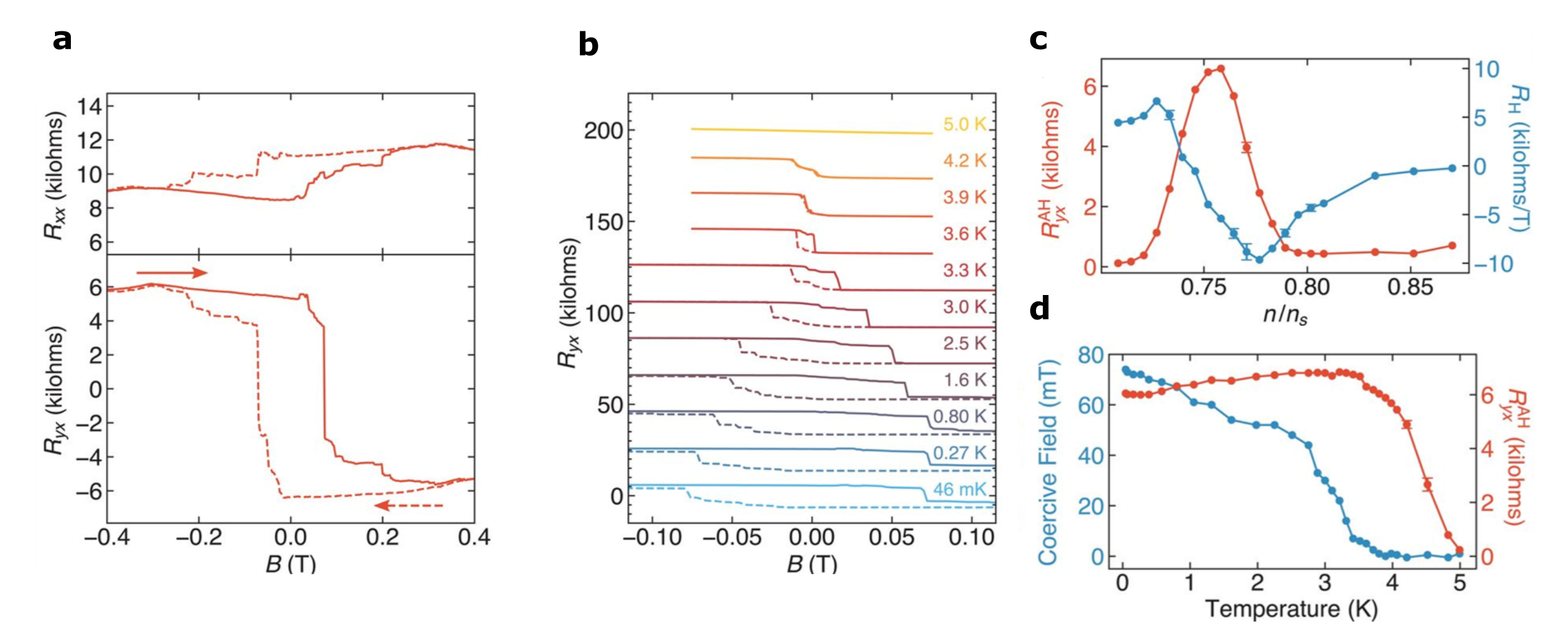}
    \vspace{-0.4cm}
    \caption{Ferromagnetic ordering near three-quarter filling. \textbf{a.} Longitudinal $R_{xx}$ (top) and (transverse) Hall resistance $R_{xy}$ (bottom) as a function of applied magnetic field, where $n/n_s = 0.746$ and $D/\epsilon_0 = \SI{-0.62}{\volt\per\nano\meter}$ at $\SI{30}{\milli\kelvin}$. The graphs reveal a hysteretic anomalous Hall effect (AHE) caused by the ferromagnetic ordering. Solid and dashed lines correspond to trace and retrace, respectively. \textbf{b.} Temperature dependence of the hysteresis loop in $R_{xy}$, which becomes gradually smaller with increasing temperature and disappears between $\SI{4.2}{\kelvin}$ and $\SI{5.0}{\kelvin}$. \textbf{c.} Zero-field AH resistance $R_{xy}^{AH}$ (red) and slope of the ordinary Hall resistance $R_H$ (blue) as a function of the relative filling factor $n/n_s$. Around three-quarter filling, $R_{xy}^{AH}$ peaks while $R_H$ changes sign. \textbf{d.} Temperature dependence of the coercive field (blue) and $R_{xy}^{AH}$ (red). Reprinted with permission from ref. \cite{ferromagnetism2019}.}
    \label{fig:AHE2}
\end{figure*}
Magnetotransport measurements on MATBG at three-quarter filling, shown in Figure \ref{fig:AHE2} revealed a significant hysteretic AHE, possibly accompanied by chiral edge states \cite{ferromagnetism2019}. 
While the ordinary Hall effect corresponds to the appearance of a transverse voltage difference across a conductor perpendicular to the electrical current and an applied magnetic field, the AHE only appears in materials with broken time-reversal symmetry, such as ferromagnets. In Figure \ref{fig:AHE2}a longitudinal $R_{xx}$ and transverse Hall resistance $R_{xy}$ are plotted as a function of applied magnetic field, at fixed filling factor $n/n_s = 0.746$, displacement field  $D/\epsilon_0 = \SI{-0.62}{\volt\per\nano\meter}$ and temperature $\SI{30}{\milli\kelvin}$. The trace and retrace graphs reveal a hysteretic AHE caused by the ferromagnetic ordering. The hysteresis loop in $R_{xy}$ becomes gradually smaller by increasing temperature and it disappears at $\SI{5.0}{\kelvin}$ as shown in Figure \ref{fig:AHE2}b. In Figure \ref{fig:AHE2}c zero-field AH resistance $R_{xy}^{AH}$ (red) and the ordinary Hall resistance $R_H$ (blue) are shown as a function of the relative filling factor $n/n_s$. Around three-quarter filling, $R_{xy}^{AH}$ peaks while $R_H$ changes sign. The temperature dependence of the coercive field (blue) and $R_{xy}^{AH}$ (red) is shown in Figure \ref{fig:AHE2}d. Time-reversal symmetry can, for instance, be broken via exchange interactions, which cause the bands in ferromagnetic materials to become exchange-split. Consequently, electrons travelling through the material are subjected to increased scattering due to spin-orbit interaction between conducting electrons and localised moments. The scattering mechanism becomes spin-dependent, which leads to asymmetric scattering due to unequal densities of spin-up and spin-down electrons. This effect \textit{anomalously} increases the transverse velocity and causes a transverse anomalous Hall current that contributes to the total Hall resistivity $\rho_{xy}$, even at zero applied field:$ \rho_{xy} = \rho_{OH} + \rho_{AH} = R_0 H + 4\pi R_A M$,
where $\rho_{OH}$ is the ordinary Hall resistivity, $\rho_{AH}$ is the anomalous Hall resistivity, $R_0$ is the Hall coefficient, $R_A$ is the anomalous Hall coefficient, $H$ is the applied magnetic field and $M$ is the magnetization of the system. Hence, the AHE can be used as a detection and characterisation technique for magnetism in a system \cite{AHE1}. For MATBG, Sharpe \textit{et al.} \cite{ferromagnetism2019} found a significantly large AHE of $\approx\SI{e4}{\ohm}$. Generally, the AHE can arise from either intrinsic or extrinsic mechanisms. The extrinsic mechanisms are related to scattering events and, in the case of an insulator, cannot contribute to the Hall resistance \cite{AHE2}. Therefore, the AHE in MATBG can only be explained by intrinsic contributions that stem from the band structure  \cite{Sinitsyn2007}.
Application-wise, the strength of this system potentially resides in its weakness. Namely weak, yet robust, magnetic ordering can easily be reversed by applying a small DC current, which can be orders of magnitudes smaller than the current required for operation in traditional devices \cite{MRAM}. Hence, this material system shows great promise for low-power electronic applications, such as magnetic memory devices \cite{ferromagnetism2019}.  
\subsection{Summary}
The manifestation of exotic phenomena in materials often hinges on electron-electron interactions. A common factor among these phenomena is the prevalence of repulsive Coulomb interactions dominating the energy landscape,  overcoming the kinetic energy of electrons which is particularly suppressed in the system with narrow, flat bands such as twisted bilayer graphene. 
In systems exhibiting correlations, like magic angle twisted bilayer graphene, these interactions give rise to emergent correlated phases, such as superconductivity, correlated insulating states and orbital ferromagnetism. A significant advantage of the twisted bilayer graphene system, distinct from the traditional correlated systems, lies in its easily tunable carrier density. This feature serves as a versatile tool, enabling the exploration of the entire phase diagram with ease and with a single device.
Flat electronic bands are at the core of the emergence of exotic quantum phenomena in correlated systems such as twisted bilayer graphene, transition metal dichalcogenide heterostructures and double bilayer graphene. 
However, these flat bands exist within a limited range of twist angles, carrier densities and temperatures, and there are still many unresolved challenges and open questions regarding these systems. Furthermore, these constraints significantly limit the accessibility of such correlated systems for both research and applications. This is further restricted by the absence of a clear and well-established theoretical framework describing this phenomena.
While substantial research efforts are imperative to deepen our understanding of the mechanisms governing the emergent phenomena in twisted 2D systems, the advent of twisted 2D materials marks a revolutionary development in condensed matter physics, with these materials, and the field of twistronics, offering an amazing playground to unveil secrets of correlated systems. 
\acknowledgments
\bibliographystyle{eplbib}
\bibliography{Collection}
\end{document}